\documentclass [12pt,preprint]{aastex}

\shorttitle{FeLoBAL Binary Quasar}
\begin{document}

\def\roma#1{\ifmmode{#1}\else{$#1$}\fi} 
\def\extra#1{\roma{\phantom{\rm#1}}}
\def\kms{\roma{\,\rm km\,s^{-1}\,}}                    % km s-1
\def\kmsmpc{\roma{\rm\,km\,s^{-1}\,Mpc^{-1}}}                       % kmsMpc
\def\Ho{\roma{\,\rm H_{o}}}                          % Ho
\newcommand{\Lya}{Ly$\alpha$}
\newcommand{\MgII}{\ion{Mg}{2}}
\newcommand{\MgI}{\ion{Mg}{1}}
\newcommand{\CaI}{\ion{Ca}{1}}
\newcommand{\CaII}{\ion{Ca}{2}}
\newcommand{\CII}{\ion{C}{2}]}
\newcommand{\OII}{[\ion{O}{2}]}
\newcommand{\CIII}{\ion{C}{3}]}
\newcommand{\CIV}{\ion{C}{4}}
\newcommand{\FeI}{\ion{Fe}{1}}
\newcommand{\FeII}{\ion{Fe}{2}}
\newcommand{\FeIII}{\ion{Fe}{3}}
\newcommand{\SiI}{\ion{Si}{1}}
\newcommand{\SiII}{\ion{Si}{2}}
\newcommand{\SiIII}{\ion{Si}{3}]}
\newcommand{\SiIV}{\ion{Si}{4}}
\newcommand{\AlIII}{\ion{Al}{3}}

\title {\bf An FeLoBAL Binary Quasar\footnote{Some of the data presented here
were obtained at the W. M. Keck Observatory, which is operated as a
scientific partnership among the California Institute of Technology,
the University of California and the National Aeronautics and Space
Administration.  The Keck Observatory was made possible by the
financial support of the W.M. Keck Foundation.}
}

\author{
Michael~D.~Gregg\altaffilmark{2,3},
Robert~H.~Becker\altaffilmark{2,3},
Richard~L.~White\altaffilmark{4}, 
Gordon~T.~Richards\altaffilmark{5},
Fred~H.~Chaffee\altaffilmark{6}, \&
Xiaohui Fan\altaffilmark{7}
}
%\email{gregg@igpp.ucllnl.org}

\altaffiltext{2}{Physics Dept., University of California, Davis, CA
95616, gregg,bob@igpp.ucllnl.org}
\altaffiltext{3}{Institute for Geophysics and Planetary Physics, L-413
Lawrence Livermore National Laboratory, 7000 East Avenue, Livermore, CA 94550}
\altaffiltext{4}{Space Telescope Science Institute, 3700 San Martin
Drive, Baltimore, MD 21218, rlw@stsci.edu}
\altaffiltext{5}{Department of Astronomy and Astrophysics, The Pennsylvania
State University, University Park, PA 16802; gtr@astro.psu.edu}
\altaffiltext{6}{W. M. Keck Observatory, 65-1120 Mamalahoa Highway,
Kamuela, HI 96743; fchaffee@keck.hawaii.edu}
\altaffiltext{7}{Institute for Advanced Study, Princeton, NJ 80540;
fan@sns.ias.edu}

\begin{abstract}

In an ongoing infrared imaging survey of quasars at Keck Observatory,
we have discovered that the $z=1.285$ quasar SDSS~J233646.2-010732.6
comprises two point sources with a separation of 1\farcs67.  Resolved
spectra show that one component is a standard quasar with a blue
continuum and broad emission lines; the other is a broad absorption
line (BAL) quasar, specifically, a BAL QSO with prominent absorption
from \MgII\ and metastable \FeII, making it a member of the
``FeLoBAL'' class.  The number of known FeLoBALs has recently grown
dramatically from a single example to more than a dozen, including a
gravitationally lensed example and the binary member presented here,
suggesting that this formerly rare object may be fairly common.
Additionally, the presence of this BAL quasar in a relatively small
separation binary adds to the growing evidence that the BAL phenomenon
is not due to viewing a normal quasar at a specific orientation, but
rather that it is an evolutionary phase in the life of many, if not
all, quasars, and is particularly associated with conditions found in
interacting systems.

\end{abstract}

\section {Introduction}

The current tally of gravitationally lensed quasars now stands at
nearly 70 (Kochanek et al.\ 2002).  The various searches which have
contributed to this total have naturally also discovered examples of
the apparently rarer phenomenon of binary quasars, of which only
$\sim 20$ systems have been documented (Mortlock, Webster, \&
Francis 1999; Kochanek et al.\ 2002).  While the gravitationally
lensed quasars have been intensively studied, binary quasars have not
received nearly as much attention, even though binary quasars may
provide invaluable insight into various aspects of the quasar
phenomenon.  Of particular importance are the cases where the pair
members exhibit strikingly different characteristics, perhaps making
it possible to deduce what aspects of the AGN environment are
responsible for presently little-understood quasar behavior such as
radio-loudness, broad or narrow absorption lines, or very red colors.
In addition, physically close binaries also supply limits,
statistically at least, on the timescales involved because once within
$\sim$ galaxy radius ($\sim 10$kpc), dynamical considerations limit
the binary lifetimes (Mortlock et al.\ 1999).

In an ongoing infrared imaging survey of quasars at Keck Observatory,
we have discovered that SDSS~J233646.2-010732.6 (hereafter
SDSS~2336-0107) is a double with a separation of 1\farcs67.  Resolved
spectra show that component~A is a standard quasar with a blue
continuum and broad emission lines.  Component~B is a broad absorption
line (BAL) quasar, specifically, a BAL QSO with prominent absorption
from \MgII\ and metastable \FeII, making it a member of the
``FeLoBAL'' class (Becker et al.\ 1997; 2000).  The number of known
FeLoBALs has increased dramatically in the last five years (Becker et
al.\ 2000; Menou et al.\ 2001; Hall et al.\ 2002), including a
gravitationally lensed example (Lacy et al.\ 2002) and the binary
discussed here, suggesting that this type of object, once thought to
be rare, may in fact be fairly common and simply overlooked in most
quasar surveys.  The presence of this BAL quasar in a relatively small
separation binary adds to the mounting evidence that the BAL
phenomenon is not simply due to viewing a normal quasar at a specific
orientation, but rather that BALs are an evolutionary phase in the
life of many, if not all, quasars, and is associated with conditions
found in interacting systems.  We adopt $\Ho = 70$ \kmsmpc,
$\Omega_{\rm m} = 0.3$, and $\Omega_{\Lambda} = 0.7$.

\section {Observations}

SDSS~2336-0107 was first identified by the Sloan Digital Sky Survey
(SDSS; York et al.\ 2000) using an early version of the quasar target
selection algorithm (Richards et al. 2002), as a quasar with z=1.285
and having modest broad absorption features (Schneider et al.\ 2002).
It was also earmarked as an ``extended'' quasar in the SDSS images,
meaning that it was slightly resolved.  Quasars at $z\gtrsim0.6$ are
unresolved in SDSS images, so the extended nature made it a target in
our imaging survey for lensed quasars at Keck Observatory.

\subsection {Infrared Imaging}

On 2001 October 30, we obtained deep imaging of SDSS~2336-0107,
totaling 350s in the $K'$ band using the Near Infrared Camera (NIRC,
Matthews \& Soifer 1994) on the Keck~I 10m telescope in 0\farcs6
seeing.  The image scale is 0\farcs15 per pixel.  The reduced image
reveals the double nature of SDSS~2336-0107 (Figure~1) and three
galaxies (g1, g2, and g3) in the field.  Photometry of the two
components was measured with {\sc daophot/iraf}\footnote{The Image
Reduction and Analysis Facilities package is distributed by NOAO,
which is operated by AURA, Inc., under contract to the National
Science Foundation}, using a bright standard star as a PSF model.  The
short exposure standard star PSF clearly differed from the PSF of the
longer exposure, dithered images of the binary quasar, so we also used
Component~B as a PSF model, even though the wings of the PSF cannot be
determined easily in this way.  In both cases, the residuals after
subtraction suggest that the shapes of two components are not quite
identical and that there is extended diffuse light at $K'$ around A,
perhaps indicating that the data are just beginning to detect the host
galaxies of one or both.  The separation of the two components is
determined to be 1\farcs673 in $K'$ and 1\farcs681 in $J$.

\subsection {Optical Imaging}

The SDSS images were obtained in relatively poor seeing, ranging from
2\farcs5 in $u^*$ to 1\farcs6 in $r^*$.  
Despite the seeing, the images prove adequate for photometry of the
two components using {\sc daophot/iraf}, a nearby bright star serving
as the PSF model and photometric zeropoint.  In $g^*r^*i^*z^*$, the PSF
fits yield relatively small photometric errors; the separations range
from 1\farcs5 to 1\farcs7, averaging $1\farcs623 \pm 0\farcs074$,
consistent to a fraction of an SDSS pixel (0\farcs396) with the
separation measured in the NIRC images, lending credibility to the
extracted photometry.  The redder Component~B is not reliably detected
in $u^*$, but an upper limited is derived from the residuals after
subtracting a single PSF at the location of component A.
Table~1 lists the optical and IR photometry, corrected for Galactic
reddening (Schlegel, Finkbeiner, \& Davis 1998).  

\subsection {Echelle Spectroscopy}

We obtained resolved spectra of the two components of SDSS~2336-0107
in 2002 January, using the Echelle Spectrograph and Imager (ESI; Epps
\& Miller 1998) on the Keck~II telescope.  The seeing was again
0\farcs6.  The ESI has a dispersion of 0.15\AA\ to 0.3\AA\ per pixel
over a wavelength range of 4000 to 10500\AA, and the 1\arcsec\ slit
used for these observations projects to 6.5 pixels.  The 900 s
exposure was obtained at an airmass of 1.33, with the slit aligned at
the position angle of the components, 95\fdg8.  The spectra show that
Component A is a standard quasar with broad emission lines while B is
found to be a BAL quasar, totally lacking any emission features
(Figure~2).  The shape of the overall spectral energy distributions
have been corrected to agree with the photometry from the SDSS images.

The redshift of component A is $z = 1.2853$ from \MgII, and $1.287$
from \CIII; as \CIII\ can be contaminated by \FeIII, \SiIII, and
\AlIII\ emission, we adopt the \MgII\ redshift.  The redshift of
component B is difficult to measure accurately because of the lack of
emission features and the broad nature of the absorption.  If we
assume that the absorption feature at 6418.8\AA, which is somewhat
broad and flat-bottomed, is due to the red half the the \MgII\ doublet
at 2803.5, the redshift is 1.2895.  There is also a narrow absorption
line at 6525.0\AA\ which, ascribed to \MgI\ 2853.0, gives $z =
1.2871$, consistent with and perhaps more accurate than the \MgII\
estimate.  Adopting this as the redshift of component~B yields an
apparent velocity difference between A and B of only $\sim 240$\kms.
Component~B also has absorption features consistent with Ca~H \& K at
a redshift of 1.2843 (Figure~2).

The ESI spectrum of component A reveals a \MgII\ $-$ \FeII\
intervening absorption system at $z = 0.8041$, perhaps due to one of
the faint galaxies which appear in the $K'$-band image (Figure~1).
There is also absorption from what may be \SiI\ 2515 and \CaI\ 2722.
The brighter galaxy (g1) is 1\farcs25 to the east of A while the
fainter (g2) is 2\farcs30 to the west, corresponding to physical
impact parameters of 9.4 and 17.4~kpc, respectively.

Also weakly detected in the spectrum of component~A is \OII\ in
emission at a wavelength corresponding to $z = 1.2882$ (Figure~2),
within 400 \kms of the emission features.  This feature is in a region
between bright sky lines and also appears on a shorter ESI exposure of
SDSS~2336-0107 taken two nights previous to the data presented here,
so its reality is beyond doubt.  Though noisy, the \OII\ feature does
appear to be broadened by a few hundred \kms, well in excess of the
instrumental resolution.

\section {Discussion}

Combining the redshifts with the SDSS and NIRC photometry, the
absolute magnitudes of the two components can be computed in the
quasar restframe as $M_{r^*}^A = -23.77$ and $M_{r^*}^B = -24.72$.
While not exceptionally bright, both components are luminous enough to
qualify as bona fide quasars.  The possibility that SDSS~2336-0107 is
gravitationally lensed is immediately ruled out by the strikingly
different spectra (Figure~2) and the short time delay between the two
paths.  The analytic formula of Witt, Mao, \& Keeton (2000) yields
time delay estimates of a few months to a year for any reasonable
lensing geometry.  The nearly identical redshifts and tiny velocity
difference, however, makes it a definite binary quasar.  For our
adopted cosmology, the apparent separation of 1\farcs67 translates to
a physical separation of $\sim 13.9$~kpc, making SDSS~2336-0107 the
smallest separation binary ever found in ground-based imaging and the
second smallest ever.  The smallest is LBQS~0103-2753 (Junkkarinen et
al.\ 2001), a 0\farcs3 (2.3~kpc) separation binary found
serendipitously during an ultraviolet spectroscopy survey of BAL
systems using HST and STIS.  In LBQS~0103-2753, the brighter component
is a standard high ionization BAL (HiBAL) quasar.

\subsection{Implications for the BAL phenomenon}

The two binary quasars with the smallest physical separations
both contain a BAL member.  Although this could be simply small number
statistics, it is natural to speculate that the BAL
features are somehow induced by the small distances between the AGN.
Mass transfer onto one of the black holes
would naturally result in a tidally interacting pair of galaxies where
the AGNs had reached relatively small separations of less than a galactic
radius.  This is further evidence supporting the view that BAL
features are a short-lived evolutionary phase during the life of a
quasar (Becker et al.\ 2000; Gregg et al.\ 2000), rather than the
result of viewing an AGN along a particular line of sight.  Additional
support of this argument comes from Canalizo \& Stockton (2001) who
show that all four of the known low ionization BAL (LoBAL) quasars
with $z < 0.4$ are in systems which exhibit tidal tails or evidence of
merging, suggesting a strong link between these phenomena and the
presence of BAL spectral features.  The orientation picture cannot so
easily account for the occurrence of BAL features in the two smallest
separation binary quasars or all four low redshift LoBAL examples,
except as by chance.

If the merger/accretion origin of BALs is correct, the material
directly responsible for the BAL spectral features is almost certainly
within tens of parsecs of the AGN and not within reach of ground-based
imaging resolution, yet mergers and tidal debris will occur over
kiloparsec scales.  It is then somewhat counter-intuitive that it is
the standard quasar component (A) of SDSS~J2336-0107 which is
sandwiched by the galaxies g1 and g2 in the NIRC image (Figure~1) and
which exhibits the narrow \OII\ in its spectrum (Figure~2).  The $K'$
image hints at a light bridge between g1 and component A, perhaps a
real physical connection.  The weak \OII\ feature of component~A is
certainly consistent with this picture, indicating some level of star
formation, quite expected for an intensely interacting system.  Some
or all of the \OII\ could come from the material in the region between
g1 and A; the 1\arcsec\ ESI slit would have entirely missed g1,
though.  The intervening \MgII\ system in the spectrum of A (Figure~2)
could arise from g2 at an intervening location; direct spectroscopy of
the galaxies in this field are required to sort out the physical
relationships.  The $K'$ seeing and S/N are not good enough to explore
the region directly between the two components more fully, but
space-based imaging would reveal details of the expected tidal
interactions which may be driving the BAL behavior.  If the two quasars are
in the nuclei of two large galaxies separated by only $\sim 13.9$~kpc,
tidal effects should be strongly pronounced.

\subsection{Implications for the FeLoBAL population}

Figure~3 plots the photometry of the two components, $u^*$ through
$K'$, highlighting the contrasting color difference.  Component~B, the
BAL, is in fact brighter in the redder bands, by $\sim 1$~magnitude in
$J$ and $K'$.  For comparison, we plot photometry synthesized from
the SDSS quasar composite (Vanden Berk, et al.\ 2001) after redshifting
to z=1.2855.  The synthetic magnitudes have been adjusted so that $r^*$
equals that of Component~A.  The excess flux in $H$ in the composite
may be due to host starlight contribution in the low redshift quasar
spectra which contribute to the composite in this wavelength region.
The even larger excess in the BAL component is probably a combination
of starlight and ``back warming'' from dust extinction.  There is
evidence that FeLoBALs are heavily reddened (Hall et al.\ 1997;
Najita, Dey, \& Brotherton 2000) and may be associated with young
galaxies and star formation (Egami et al.\ 1996).

Perhaps the most intriguing finding is that Component~B is an FeLoBAL,
a quasar with low ioniziation broad absorption from \MgII\ plus broad
and narrow absorption from metastable states of \FeII.  In Figure~4,
the spectrum of Component~B is compared to two other FeLoBALs,
J1556+3517 (Becker et al.\ 1997) and the original of the class,
Q0059-2735 (Hazard et al.\ 1987).  The spectrum of SDSS~2336-0107~B
exhibits strong \FeII\ at $\lambda\lambda 2380, 2600,$ and 2750\AA.
The broad \MgII\ absorption depresses the spectrum blueward to at
least 2630\AA, or -18,000 \kms in the quasar restframe.  While not
quite as red as J1556+3517, SDSS~2336-0107~B has essentially no
emission lines whatsoever.

There is now one FeLoBAL quasar (SDSS~2336-0107) known among the 15
binary QSO systems and one FeLoBAL (J1004+1229; Lacy et al.\ 2002)
among the $\sim 70$ known gravitationally lensed quasars.  These
statistics argue strongly that FeLoBALs must be much more common than
their presently cataloged population suggests.  Lacy et al.\ (2002)
discuss the details of red quasar luminosity functions and lensing
probabilities, but not with regard to FeLoBALs in particular.  With 1
FeLoBAL in $\sim 70$ lensed quasars and one among 40 quasars in $\sim
20$ binary systems, the simplest explanation is that from 1\% to a few
percent of quasars are FeLoBALs, or alternatively, that the FeLoBALs
phase lasts for a similar percentage of the total quasar lifetime.
The very red spectral energy distribution and weak or completely
absent broad emission lines (Figure~1) makes these objects extremely
difficult to find, especially at redshifts $\gtrsim 2.5$ where the
\MgII\ absorption edge moves into the $J$ band.  An IR-selected sample
of quasars is needed to better gauge the frequency of FeLoBAL numbers,
but evidence is mounting (Becker et al.\ 1997; Lacy et al.\ 2002; Hall
et al.\ 2002) that they are much more common than suspected even just
a few years ago when the rather blue example of Q0059-2735 was the
only one known (Hazard et al.\ 1987).

\section{Conclusions}

A deep imaging survey of BAL quasars and a control sample of non-BAL
quasars, from the ground or space, is needed to document the frequency
of each in systems that show tidal interactions, which would be a more
comprehensive version of the work done by Canalizo \& Stockton (2001).
If BAL-ness is explained simply by viewing an ordinary quasar along a
line of sight which skims the ``dusty torus'' (Weymann et al.\ 1991),
then the two populations should appear with equal frequency in merging
or interacting systems.  If BAL features are produced by conditions
created in mergers, and orientation plays less or no role, then they
will be more common in chaotic systems with tidal tails and other
signs of host galaxy interactions.  Dynamical modeling of the tidal
effects may help constrain the timescales of various quasar phases.

FeLoBALs may be a common feature of the AGN landscape, especially if
there are Seyfert-type luminosity class analogs to the brighter
objects now being unearthed.  If FeLoBALs are indeed common, then
infrared sky surveys which reach deeper than 2MASS, such as those
being undertaken with SIRTF, will turn up numerous examples along with
other extreme BAL phenomena in quasar and galaxy luminosity objects.
The FeLoBAL-like Hawaii~167 (Cowie et al.\ 1994) was discovered in
just such a survey.  Such surveys are needed to provide more accurate
estimates of the extremely red quasar population, especially at
redshifts above 2.5 where such objects are essentially invisible in
the optical.

\acknowledgments

We thank Mark Lacy for helpful discussions.
The authors wish to recognize and acknowledge the very significant
cultural role and reverence that the summit of Mauna Kea has always
had within the indigenous Hawaiian community.  We are most fortunate
to have the opportunity to conduct observations from this mountain.
Part of the work reported here
was done at the Institute of Geophysics and Planetary Physics, under
the auspices of the U.S. Department of Energy by Lawrence Livermore
National Laboratory under contract No.~W-7405-Eng-48.  
Funding for the creation and distribution of the SDSS Archive has been
provided by the Alfred P.  Sloan Foundation, the Participating
Institutions, the National Aeronautics and Space Administration, the
National Science Foundation, the U.S. Department of Energy, the
Japanese Monbukagakusho, and the Max Planck Society. The SDSS Web site
is http://www.sdss.org/.

%%%%%%%%%%%%%%%%%%

\pagebreak

\begin {deluxetable}{llllllrrrrrrl}
\tabletypesize{\scriptsize}
\tablewidth{0in}
\tablecaption{Binary Quasar Details}
%\tablehead{\multicolumn{2}{c}{Red Quasar Properties}\\
\tablehead{
\colhead{Name} &
%\colhead{$\alpha\tablenotemark{a}$ (J2000) } &
%\colhead{$\delta\tablenotemark{a}$ } &
\colhead{$\alpha$ (J2000) } &
\colhead{$\delta$ } &
\colhead{Gal.\ $A_V$} &
\colhead{$z$} &
\colhead{$u^*$} &
\colhead{$g^*$} &
\colhead{$r^*$} &
\colhead{$i^*$} &
\colhead{$z^*$}  &
\colhead{$J$\tablenotemark{a}} &
\colhead{$K$\tablenotemark{a}} & 
\colhead{$M_{r^*}$\tablenotemark{b}} 
}
\startdata
Component A &  
$\mathrm{ 23^{h} 36^{m} 46\fs2 }$  &
 $-01\arcdeg 07\arcmin 32\farcs6$  &
 0.114 &
 1.2853 &
\phn\phn 19.75  & 
 19.65  & 
 19.35   & 
 19.26   & 
 19.12  & 
 18.19  & 
 17.22  & 
 -23.77 \\
Component B &  
$\mathrm{ 23^{h} 36^{m} 45\fs1 }$  &
 $-01\arcdeg 07\arcmin 32\farcs4$   &
 0.114 &
 1.2871 &
 $>22.2\phn$  & 
 21.37   & 
 19.71   & 
 18.94   & 
 18.68   & 
 17.24   & 
 16.23   &
 -24.72  \\
\enddata
\tablenotetext{a} {J and K photometry have been corrected to the AB
magnitude system using zeropoint offsets of 0.908 and 1.842
magnitudes, determined from a model of Vega.}
\tablenotetext{b} {Computed in quasar rest frame}
%\tablenotetext{b} {Optical magnitudes measured from flux calibrated spectra}
%\tablenotetext{c} {2MASS catalog values}
%
%\tablecomments
%Quantities in parentheses corrected for intrinsic
%extinction of A$_{\rm B}=1.1$}
\end {deluxetable}

\pagebreak

\begin{figure}
\epsscale{0.7}
\plotone{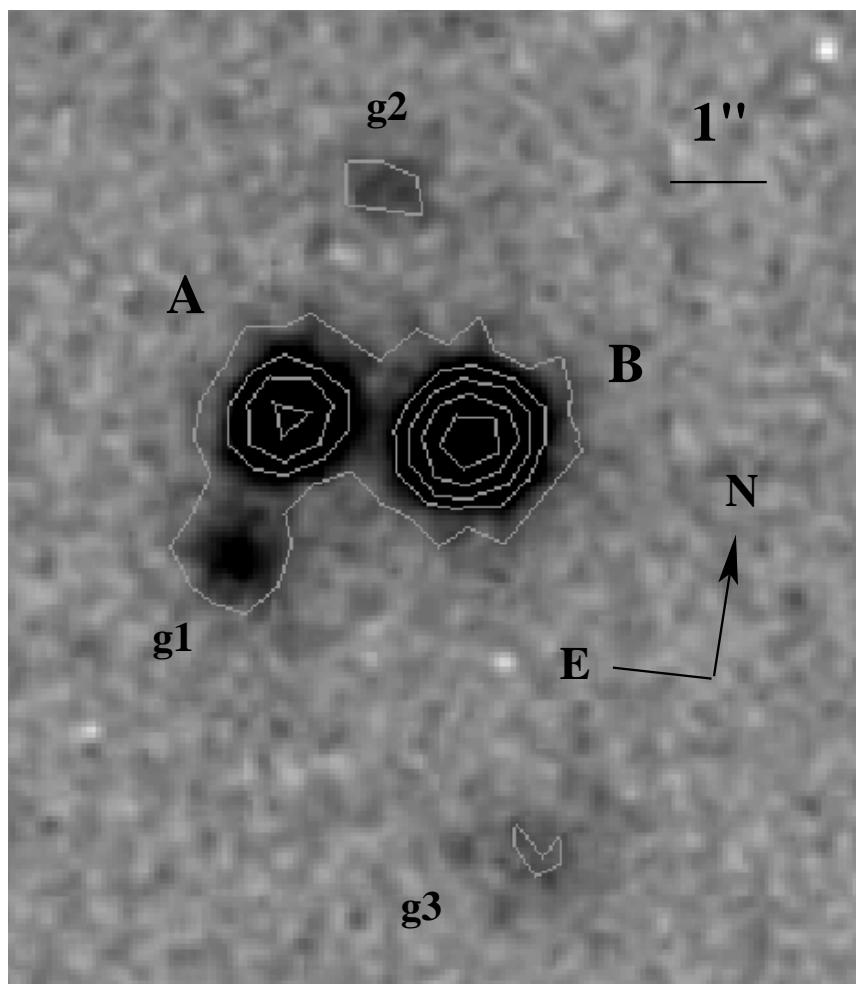}
\caption{The Keck NIRC $K'$ image of SDSS~2336-0107, obtained in
seeing of 0\farcs6; total exposure is 350s.  Components A and B are
separated by 1\farcs67.  Component B is 1.0 magnitude brighter than A
at $2\mu m$.  The excess light between galaxy g1 and A may indicate a
physical connection. }
\end{figure}

\begin{figure}
\epsscale{1.}
%\plotone{ABplot2.ps}
\plotone{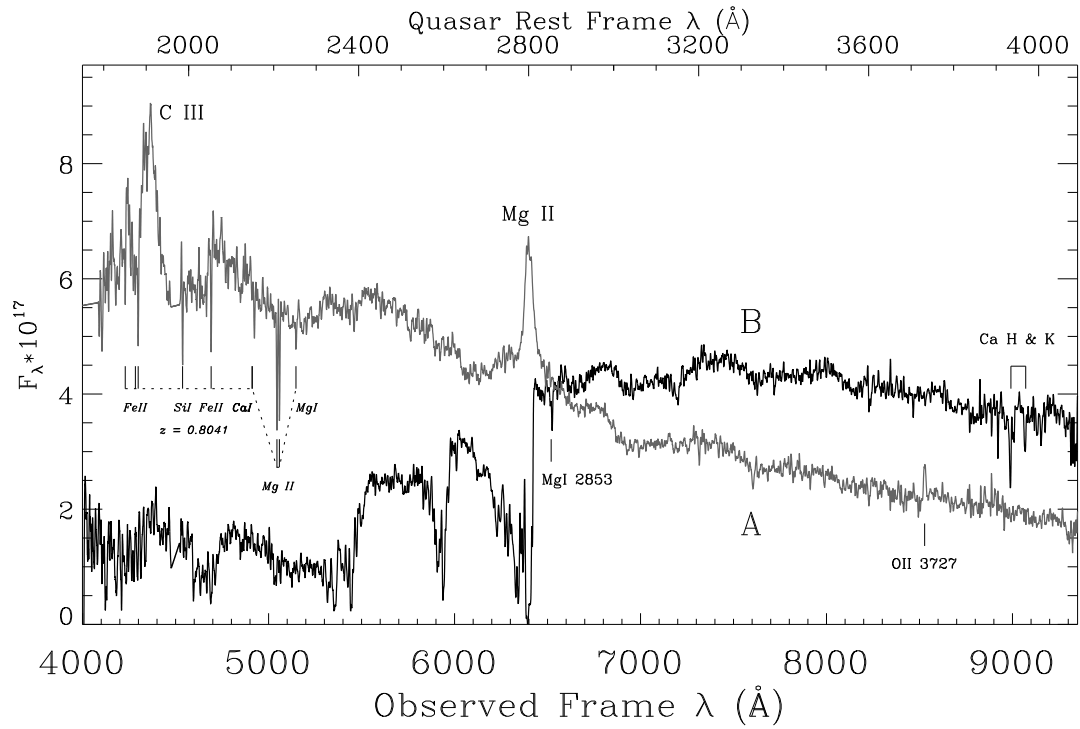}
\caption{Keck ESI spectra of the two components of SDSS~2336-0107
showing the striking difference in spectral appearance of the two
components; A is a typical quasar with broad emission lines and blue
continuum while B exhibits no emission lines whatsoever and has a
relatively red spectal energy distribution.}
\end{figure}

\begin{figure}
\epsscale{0.7}
%\plotone{photplotab.ps}
\plotone{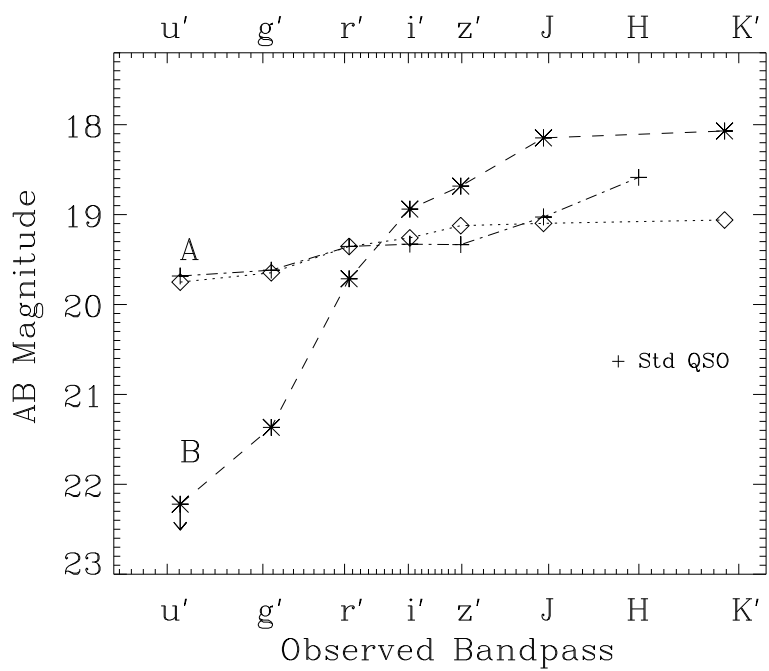}
\caption{Comparative photometry of the two components showing the
similarity of component A (open diamonds, dotted line) to a typical
quasar as represented by the SDSS composite spectrum (plus signs,
dash-dot line) and highlighting the red spectral energy distribution
of B (asterisks, dashed line).  The $u^*g^*r^*i^*z^*$ photometry was
measured from the SDSS image of this field using PSF fitting in {\sc
daophot/iraf}.  The $J$ and $K'$ photometry from Keck/NIRC has been
shifted to the AB system using constants of 0.908 and 1.842 magnitudes
respectively.}
\end{figure}

\begin{figure}
\epsscale{1.0}
%\plotone{felobalcomp2.ps}
\plotone{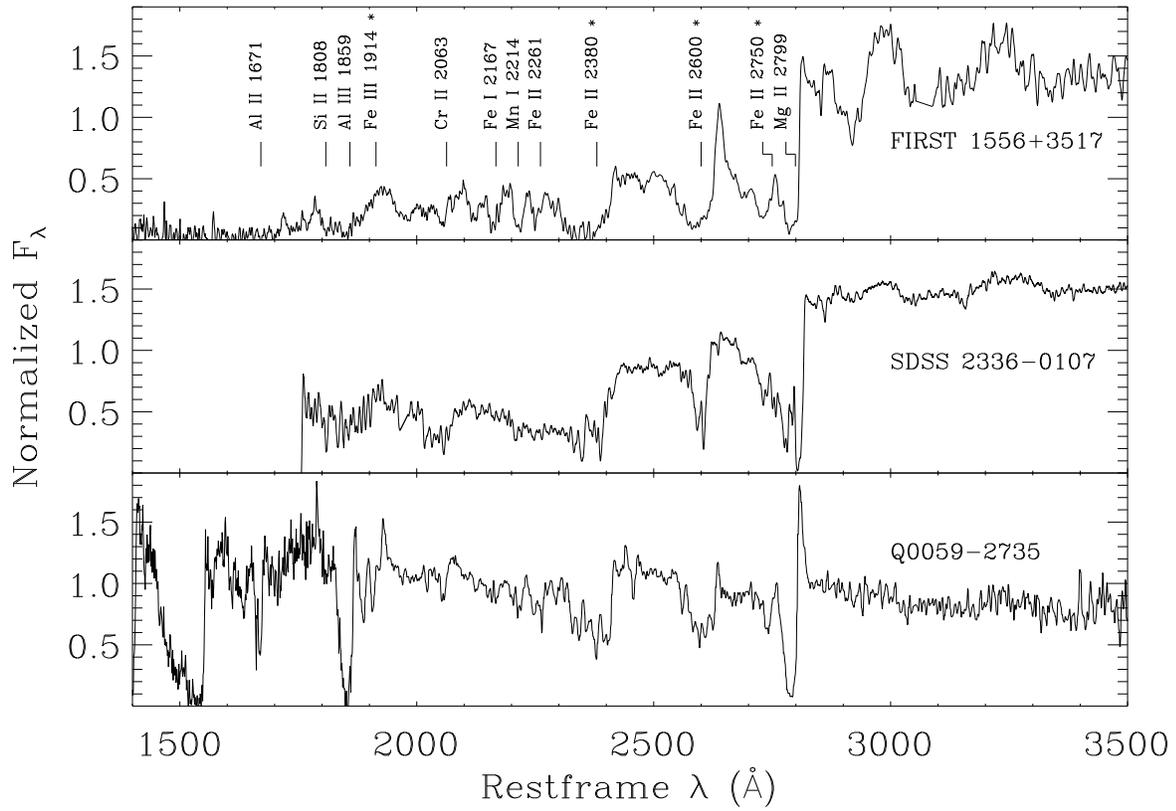}
\caption{Spectrum of the FeLoBAL member of SDSS~2336-0107 (B) compared
to FIRST~J1556+3517 and Q0059-2735.  Locations of prominent absorption
features are marked; an asterisk indicates a metastable transition.}
\end{figure}
\end{document}